\documentclass[11pt]{article}

\usepackage{color}
\usepackage{graphicx}
\usepackage{graphics}
\usepackage{amscd}
\usepackage{amsmath}
\usepackage{amsfonts}
\usepackage{amssymb}
\usepackage{cite}
\usepackage{pifont}

\newcommand{\blue}{\color{blue}}

\newcommand{\bF}{ {\mathbb F}}

\newcommand{\C}{ {\mathcal C}}

\newtheorem{theorem}{Theorem}[section]

\newtheorem{conjecture}[theorem]{Conjecture}
\newtheorem{corollary}[theorem]{Corollary}

\newtheorem{definition}[theorem]{Definition}
\newtheorem{example}[theorem]{Example}

\newtheorem{lemma}[theorem]{Lemma}

\newtheorem{proposition}[theorem]{Proposition}
\newtheorem{remark}[theorem]{Remark}

\newcommand{\EOP} { \hfill $\Box$ }

\begin{document}

\title{Strict Half-Singleton Bound, Strict Direct Upper Bound for Linear Insertion-Deletion Codes and Optimal Codes}

\author{Qinqin Ji, Dabin Zheng, Hao Chen and Xiaoqiang Wang
  \thanks{ Qinqin Ji, Dabin Zheng and Xiaoqiang Wang are with the Hubei Key Laboratory of Applied Mathematics, Faculty of Mathematics and Statistics, Hubei University, Wuhan 430062, China, E-mail: qqinji@163.com, dzheng@hubu.edu.cn, waxiqq@163.com. Hao Chen is with the College of Information Science and Technology/Cyber Security, Jinan University, Guangzhou, Guangdong Province, 510632, China, E-mail: haochen@jnu.edu.cn. The corresponding author is Dabin Zheng. The research of Dabin Zheng and Xiaoqiang Wang was supported by NSFC Grant 11971156 and NSFC Grant 12001175.  The research of Hao Chen was supported by NSFC Grant 62032009 and Major Program of Guangdong Basic and Applied Research under Grant 2019B03032008.}}

\date{ }

\maketitle

\begin{abstract}

Insertion-deletion codes (insdel codes for short) are used for correcting synchronization errors in communications, and in other many interesting fields such as DNA storage, date analysis, race-track memory error correction and language processing, and have recently gained a lot of attention. To determine the insdel distances of linear codes is a very challenging problem. The half-Singleton bound on the insdel distances of linear codes due to Cheng-Guruswami-Haeupler-Li is a basic upper bound on the insertion-deletion error-correcting capabilities of linear codes. On the other hand the natural direct upper bound $d_I(\C) \leq 2d_H(\C)$ is valid for any insdel code. In this paper, for a linear insdel code $\C$ we propose a strict half-Singleton upper bound $d_I(\C) \leq 2(n-2k+1)$ if $\C$ does not contain the codeword with all 1s, and a stronger direct upper bound $d_I(\C) \leq 2(d_H(\C)-t)$ under a weak condition, where $t\geq 1$ is a positive integer determined by the generator matrix. We also give optimal linear insdel codes attaining our strict half-Singleton bound and direct upper bound, and show that the code length of optimal binary linear insdel codes with respect to the (strict) half-Singleton bound is about twice the dimension. Interestingly explicit optimal linear insdel codes attaining the (strict) half-Singleton bound, with the code length being independent of the finite field size, are given.

\vspace{3mm}

\end{abstract}

\section{Introduction }

In most communication and storage channels, the most common type of errors are substitution errors, in which a transmitted symbol is replaced with another symbol.
However, channels may also suffer from synchronization errors due to slips in synchronization causing the deletion of a symbol from a message or the insertion of an extra symbol into a message~\cite{Golomb1963, Tanaka1976}. Insdel codes were introduced in~\cite{Levenshtein1965} for correcting synchronization errors. Insdel errors model also has been widely applied in many interesting fields such as DNA storage, date analysis, race-track memory error correction and language processing, we refer to \cite{Brill2000,Chvatal1975, Cheng2018,SWGY2017,Chee2017,Haeupler2017,Haeupler2018,Haeupler2019,GS2019,Jain2017,Lenz2020,Sankoff1983,Och2003,Xu2005} for the construction and application of insdel codes.

For a vector ${\bf a} \in \bF_q^n$,  the support of ${\bf a}$ is  ${\rm supp}({\bf a}) =\{i\, :\, a_{i} \neq 0\}$. The Hamming weight $w_H({\bf a})$ of ${\bf a}$ is the number of coordinate positions in its support. The Hamming distance $d_H({\bf a}, {\bf b})$ between two vectors ${\bf a}$ and ${\bf b}$ is defined to be the Hamming weight $w_H({\bf a}-{\bf b})$. For a linear code $\C \subset \bF_q^n$ of dimension $k$, its Hamming distance (or weight) $d_H$ is the minimum of Hamming distances $d_H({\bf a}, {\bf b})$ between any two different codewords ${\bf a}$ and ${\bf b}$ in $\C$. It is well-known that the Hamming distance (or weight) of a linear code $\C$ is the minimum Hamming weight of non-zero codewords. The famous Singleton bound $d_H \leq n-k+1$ is the basic upper bound for linear error-correcting codes. For two codewords ${\bf a}  \neq {\bf  b}$ in a code $\C \subset \bF_q^n$, the insdel distance $d_I({\bf a}, {\bf b})$ between them is defined as the smallest number of insertions and deletions needed to transform one codeword into the other. Similarly to the minimum Hamming distance, the minimum insdel distance of a code is defined as the minimum insdel distance among all its distinct codewords. It is easy to verify that the insdel distance is a metric, and a code or a linear code is called a insdel code or a linear insdel code if we consider insdel metric. The minimal insdel distance of an insdel code is an important parameter, which determines its insertion-deletion error-correcting capability.

The study of insdel codes dates back to the pioneering work of Levenshtein~\cite{Levenshtein1965}. From then on, the problem to correct the synchronization errors has attracted lots of continuous efforts. For the recent progress in insdel codes, the reader can refer to the nice survey~\cite{Haeupler2021} and references therein. Since linear codes have a compact representation, and are efficiently encodable (decodable), we recall the main research progress in linear insdel codes below. In 2010, Abdel-Ghaffar et al.~\cite{Abdel2010} showed that an $[n,k]$ linear code $\C$ over $\mathbb{F}_q$ with $n<2k$ had the minimum insdel distance $d_I(\C)=2$, and gave a sufficient and necessary condition for $d_{I}(\C)=2$. Actually it was shown in ~\cite{Haeupler2017} that
$$d_{I}({\bf a}, {\bf b})=2(n-\ell),$$
where $\ell$ is the length of a longest common subsequence of ${\bf a}$ and ${\bf b}$. It is clear $d_I({\bf a}, {\bf b}) \leq 2d_H({\bf a}, {\bf b})$ since  $\ell \geq n-d_H({\bf a}, {\bf b})$ is valid for arbitrary two different vectors ${\bf a}$ and ${\bf b}$ in $\bF_q^n$. We call the natural upper bound $d_I(\C) \leq 2d_H(\C)$ the direct upper bound for insdel codes. It is true for any insdel code, not only linear insdel codes. Hence it was shown in Haeupler et al.~\cite{Haeupler2017} that the minimum insdel distance $d_I(\C)\leq 2(n-k+1)$ for any $[n,k]$ linear code $\C$ over $\mathbb{F}_q$ from the Singleton bound on the Hamming distances, which is called the direct Singleton bound for linear insdel codes. For insertion-deletion codes the ordering of coordinate positions strongly affects the insdel distances. In this paper we give some upper bounds for insdel distances of linear codes which are valid for any fixed ordering of coordinate positions. There have been many constructions of insdel codes in previous works \cite{Tonien2007,Wang2004,MacAven2007,Con2021,Duc2021,Cheng2021,Chen2022,Abdel2010,GS2019,SWGY2017}.

In~\cite{Duc2021}, Do Duc et al. showed that the minimum insdel distance of any $[n, k]$ Reed-Solomon (RS) code over $\bF_q$ is no more than $2n-2k$ if $q$ is large enough compared to the code length $n$. Then, Chen et al.~\cite{Chen2022} generalized this result and showed that for any $[n,k]$ linear code over $\bF_q$ with $n> k\geq 2$, the minimum insdel distance is at most $2n-2k$, and an infinite family of optimal two-dimensional RS codes meeting the bound was constructed. Very recently, Cheng, Gruswami, Haeupler and Li ~\cite{Cheng2021} proved the existence of binary linear codes of length $n$ and rate just below $\frac{1}{2}$
capable of correcting $\Omega(n)$ insertions and deletions, and proposed the asymptotic half-Singleton bound for the insdel distances of an $[n, k]$ linear code over
$\bF_q$, then their results were improved significantly in~\cite{Con2021}. Their half-Singleton bound for an $[n,k]$ linear code $\C$ over $\mathbb{F}_q$ can be reformulated as $d_I(\C) \leq \max\{ 2(n-2k+2), 2\}$. For a simpler proof we refer to ~\cite{Chen2021}. A new coordinate-ordering free upper bound was also given in ~\cite{Chen2021}. It is well-known that the half-Singleton bound is only true for linear insdel codes.

When the dimension $k=2$, RS codes attaining the bound $2n-4$ were constructed in~\cite{Chen2022,Cheng2021,Con2021}. When the dimension $k\geq 3$, the half-Singleton bound is tighter than $2n-2k$. Con et al.~\cite{Con2021} proved that there were $[n, k]$ RS codes achieving the half-Singleton bound if the field size was large enough and gave a deterministic construction of such codes over much larger fields (of size $n^{k^{O(k)}}$). The code length is small when compared to the field size. As far as we known, up to now there is no explicit construction of optimal $[n, k]$ linear insdel codes attaining the half-Singleton bound for $k\geq 3$.

The direct upper bound $d_I(\C) \leq 2d_H(\C)$ is fundamental for insdel codes and the half-Singleton upper bound is fundamental for linear insdel codes. When $d_H \leq n-2k+1$, the direct upper bound has to be used to upper bound the insdel distances of codes. In this paper, we show both upper bounds for linear codes can be improved under a weak condition. We first propose a strict half-Singleton upper bound
$$d_I(\C) \leq 2(n-2k+1)$$
on the insdel distance for linear insdel codes without the codeword with all 1s by investigating the linear equations associated with the generator matrices. Then, we provide a sufficient condition for a linear insdel code attaining the strict half-Singleton bound, and by this sufficient condition some optimal linear insdel codes with dimension $k\geq 3$ are constructed. Finally, we study the optimal binary linear insdel codes with respect to the (strict) half-Singleton bound and prove that the code length of optimal binary linear insdel codes is about twice the dimension, and conjecture that optimal binary linear insdel codes have parameters $[2k, k, 4]$ or $[2k+1, k, 4]$ with respect to the half-Singleton bound or the strict half-Singleton bound, respectively. Interestingly explicit optimal linear insdel codes attaining the (strict) half-Singleton bound, with the code length being independent of the finite field size, are obtained. On the other hand we prove that the direct upper bound $d_I(\C) \leq 2d_H(\C)$ for arbitrary insdel codes can be improved to $d_I(\C) \leq 2(d_H(\C)-t)$ for linear insdel codes, where $t\geq 1$ is a positive integer determined by the generator matrix. Some examples attaining our strict direct upper bound are given.

The rest of this paper is organized as follows. In section 2, we introduce some definitions of insdel codes and preliminary results on linear insdel codes. Section~3 proposes the strict half-Singleton bound for linear insdel codes and gives another proof of the known half-Singleton bound. In section~4, we give a sufficient condition for constructing optimal linear insdel codes with respect to our strict half-Singleton bound and provide some examples of optimal linear insdel codes. In section~5, we study optimal binary linear insdel codes and show that the code length of optimal binary linear insdel codes is about twice the dimension.  In Section 6, we prove the strict direct upper bound and discuss the optimal linear insdel codes attaining this bound. Finally, Section~7 concludes this paper.

\section{Preliminaries }

Let $\bF_q$ be a finite field of $q$ elements and $\bF_q^n$ be a vector space over $\bF_q$ with dimension $n$. A subspace $\C$ of $\bF_q^n$ over $\bF_q$ is called a linear code of  length $n$ over $\bF_q$.
Its dual $\C^{\perp}$ is a linear code $\C^{\perp}=\{(x_1, x_2 \cdots, x_n) \in \bF_q^n\,:\, \sum_{i} x_i y_i=0, \forall (y_1, \ldots, y_n) \in \C\}$. As mentioned above, the Hamming distance of a linear code
equals the minimum Hamming weight of its non-zero codewords. A linear code $\C$ is called projective if $d_H(\C^{\perp}) \geq 3$. That is, any two columns of the generator matrix of $\C$ are linear independent
over $\bF_q$. The following result shows the property of columns of the generator matrix of a linear code.

\begin{lemma}\label{lemtr}
Let $\C$ be an $[n,k]$ linear code over $\mathbb{F}_{q}$ and denote ${d}_{H}(\C)$ the minimal Hamming distance of $\C$. Let $G$ denote the generator matrix of $\C$. Then there are $k$ of any $n-{d}_{H}(\C)+1$ columns of ${G}$ are linearly independent over $\mathbb{F}_{q}$.
\end{lemma}

{\it Proof.}  Let $s=n-{d}_{H}(\C)+1$ and $G=(G_1,G_2,\cdots,G_n)$, where $G_i=(g_{1i},g_{2i},\cdots,g_{ki})^T$. Assume that there exist $s$ columns $G_{j_1},G_{j_2},\cdots,G_{j_s}$ of $G$ such that any $k$ vectors of them are linearly dependent over $\mathbb{F}_q$, that is the rank of the matrix $\bar{G}=(G_{j_1},G_{j_2},\cdots,G_{j_s})$ is less than $k$. So, the linear system
\begin{equation}\label{fdzero}
(x_1,x_2,\cdots,x_k)\bar{G}={\bf 0},
\end{equation}
has nonzero solutions. Let $\mathbf{y}=(y_1,y_2,\cdots,y_k)\in \bF_q^k$ be a nonzero solution of (\ref{fdzero}) and $\mathbf{c}=\mathbf{y}G$. Then $w_{H}(\mathbf{c})\leq n-s=d_H(\C)-1$. This is a contradiction.                         \EOP

In this paper, we mainly consider the insdel distance of linear codes used in high insertions and deletions noise regime. We give the definition of the insdel distance of two vectors as follows.
\begin{definition}
For two vectors $\mathbf{a},\mathbf{b}\in \mathbb{F}_q^n$, the insdel distance $d_{I}(\mathbf{a},\mathbf{b})$ between $\mathbf{a}$ and $\mathbf{b}$ is the minimal number of insertions and deletions which are needed to transform $\mathbf{a}$ into $\mathbf{b}$. It can be verified that $d_I(\mathbf{a},\mathbf{b})$ is indeed a metric on $\mathbb{F}_q^n$.
\end{definition}

Let $\mathbf{a} = (a_1, a_2, \cdots, a_n), \mathbf{b} = (b_1, b_2, \cdots, b_n)\in \bF_q^n$ be two sequences (or vectors). A common subsequence of $\mathbf{a}$ and $\mathbf{b}$ is a sequence $(c_1, c_2, \cdots, c_m)$ such that
$c_s= a_{i_s} = b_{j_s}$ for $1\leq s\leq m$, $1 \leq  i_1 < i_2 < \cdots < i_m\leq n$ and $1 \leq j_1 < j_2 < \cdots <j_m \leq n$. It has been proved that the insdel distance between any two vectors can be
characterized by their longest common subsequences.

\begin{lemma}\cite[Lemma 1]{Duc2021}\label{lemlcs}
Let $\mathbf{a},\mathbf{b}\in \mathbb{F}_q^n$. Then we have$$d_I(\mathbf{a},\mathbf{b})=2n-2\ell(\mathbf{a},\mathbf{b}),$$ where $\ell(\mathbf{a},\mathbf{b})$ denotes the length of a longest common subsequence of $\mathbf{a}$ and $\mathbf{b}$.
\end{lemma}

For any two distinct codewords $\mathbf{a}, \mathbf{b}\in \C$, by Lemma~\ref{lemlcs} we know that $d_I(\mathbf{a},\mathbf{b})$ is even and $d_I(\mathbf{a},\mathbf{b})\geq 2$.
Like the Hamming distance, the insdel distance of a linear insdel code $\C$ over $\mathbb{F}_q$ is defined as
\begin{equation}\notag
d_I(\C)=\min_{\mathbf{a},\mathbf{b}\in \C, \mathbf{a}\neq \mathbf{b}}\left\{\,\,  d_I( \mathbf{a},\mathbf{b}) \,\, \right\}.
\end{equation}

A linear code $\C$ over $\bF_q$ of length $n$, dimension $k$ and minimum insdel distance $d_I(\C)$ is called an $[n, k, d_I(\C)]$ linear insdel code over $\bF_q$. Like
the Hamming metric, an $[n, k, d_I]$ linear insdel code $\C$ has insdel error-correcting capability up to $\lfloor\frac{d_I-1}{2}\rfloor$~\cite{Duc2021}.
So, an $[n, k, d_I]$ linear code $\C$ can correct insdel errors if and only if $d_I>2$. As mentioned in the first section, Chen et al. generalized the Singleton type bound of linear insdel codes~\cite{Haeupler2017} to the following case.

\begin{lemma}\cite[Theorem A]{Chen2022}\label{lemboc}
Let $\C$ be an $[n,k]$ linear code over $\mathbb{F}_q$ with $n>k>2$. Then the minimum insdel distance of $\C$ is at most $2n-2k$, i.e., $d_I(\C)\leq 2n-2k$.
\end{lemma}

Cheng, Gruswami, Haeupler and Li proposed the half-Singleton bound for linear insdel codes in ~\cite{Cheng2021}. The non-asymptotic version of half-Singleton bound and a simple proof was given in ~\cite{Chen2021}.

\begin{lemma}[Half-Singleton bound \cite{Cheng2021}]\label{lemChenbound}
Let $\C$ be a non-degenerate linear $[n,k]$ code over $\bF_q$. Its insdel distance satisfies
\begin{equation}\notag
d_I(\C) \leq \max\left\{  2(n-2k+2),\, 2 \right\}.
\end{equation}
\end{lemma}

The following lemma shows that a linear insdel code must contain two special codewords if its minimal insdel distance  is equal to $2$.

\begin{lemma}\cite[Lemma~1]{Abdel2010}\label{lemsec}
Let $\C$ be an $[n,k]$ linear code over $\mathbb{F}_q$. Then, $d_I(\C)=2$ if and only if $\C$ contains a codeword $\mathbf{c}=(c_1,c_2,\cdots,c_n)$ such that, for some $1\leq u\leq v\leq n$ and $\alpha \in \mathbb{F}_q$, $\mathbf{x}=(x_1,x_2,\cdots,x_n)$ defined by
$$ x_i=\left\{
\begin{array}{lrl}
0,          &  & for\ 1\leq i<u \ or\ v<i\leq n\\
c_{i+1}-c_i,&  & for\ u\leq i<v\\
\alpha,          &  & for\ i=v
\end{array}
\right.
$$
is a nonzero codeword.
\end{lemma}

\section{Strict half-Singleton bound}

In this section, we show that the half-Singleton bound on the insdel distance of a linear code $\C$ can be improved if $\mathbf{1} \notin \C$. Based on this improved upper bound, we give another proof of the half-Singleton bound on the minimal insdel distance of linear insdel codes and some useful corollaries. To this end, we first introduce some notation. For positive integers $n$ and $s$ with $s\leq n$, we denote $[n]=\{1, 2, \cdots, n\}$ and $[n]^s$ the set of all vectors of length $s$ whose coordinates are from $[n]$. We say a vector $I=(I_1, I_2, \cdots, I_s)\in [n]^s$ is an increasing vector if its coordinates are monotonically increasing, i.e., for any $u<v$ we have $I_u<I_v$, where $I_u$ is the $u$th coordinate of~$I$.

\begin{theorem}[Strict half-Singleton bound]\label{thm:upperbound}
Let $\C$ be an $[n,k]$ linear code over $\mathbb{F}_{q}$. If $\mathbf{1}=(1,1,\cdots,1)\notin \C$, then the insdel distance of $\C$ satisfies
\begin{equation}\notag
d_{I}(\C) \leq  \max \left\{\, 2(n-2k+1), \, 2 \, \right\}.
\end{equation}
\end{theorem}

{\it Proof:} Let $G$ denote a generator matrix of $\C$ and $d$ denote the minimum Hamming distance of $\C$. Then there exists a codeword $\mathbf{c}\in \C$ satisfying $w_H(\mathbf{c})=d$. Thus, $\ell(\mathbf{0},\mathbf{c})=n-d$.

Next, we discuss the insdel distance of $\C$ in two cases.

(1) $n-d\geq 2k-1$. Then, $\ell(\mathbf{0},\mathbf{c})=n-d$ and
 \begin{equation*}
 d_I(\C)\leq d_I(\mathbf{0},\mathbf{c})=2(n-\ell(\mathbf{0},\mathbf{c}))\leq2(n-2k+1).
 \end{equation*}

(2) $n-d< 2k-1$. In the following we show that there exist two different codewords $\mathbf{a},\mathbf{b}\in \C$ such that $\ell(\mathbf{a},\mathbf{b})=n-1$ or $\ell(\mathbf{a},\mathbf{b})\geq 2k-1$. This implies that
\begin{equation*}
 d_I(\C)\leq d_I(\mathbf{a},\mathbf{b})=2(n-\ell(\mathbf{a},\mathbf{b}))\leq \max\{2(n-2k+1),2\}.
 \end{equation*}

 Let $G=(G_1,G_2,\cdots,G_n)$ be a generator matrix of $\C$, where $G_i$ denote the $i$th column of $G$. Let $I=(I_1,I_2,\cdots,I_s)$ and $J=(J_1,J_2,\cdots, J_s)$ be two increasing vectors of $[n]^s$. Define an $2k\times s$ matrix as follows:
 \begin{equation}\label{fdfir}
 M_{IJ}=\begin{pmatrix}G_{I_1} & G_{I_2} & \cdots & G_{I_s}\\
 G_{J_1} & G_{J_2} & \cdots & G_{J_s}\end{pmatrix}.
 \end{equation}
 Consider the linear equations
 \begin{equation}\label{fdsec}
  (\mathbf{x},-\mathbf{y})M_{IJ}={\bf 0},
 \end{equation}
where $\mathbf{x}=(x_1,x_2,\cdots,x_k)$ and $\mathbf{y}=(y_1,y_2,\cdots,y_k)$. If the system (\ref{fdsec}) has a nonzero solution $(\mathbf{x},-\mathbf{y})\in \mathbb{F}_q^{2k}$ with $\mathbf{x}\neq \mathbf{y}$, then there exist two codewords $\mathbf{a}=(f_1(\mathbf{x}),f_2(\mathbf{x}),\cdots,f_n(\mathbf{x}))$ and $\mathbf{b}=(f_1(\mathbf{y}),f_2(\mathbf{y}),\cdots,f_n(\mathbf{y}))$ in $\C$, where $f_i(\mathbf{x})=\mathbf{x}G_i$, $f_i(\mathbf{y})=\mathbf{y}G_i$ for $i=1,2,\cdots,n$ such that
\begin{equation}\notag
 (f_{I_1}(\mathbf{x}),f_{I_2}(\mathbf{x}),\cdots,f_{I_s}(\mathbf{x}))=(f_{J_1}(\mathbf{y}),f_{J_2}(\mathbf{y}),\cdots,f_{J_s}(\mathbf{y})).
\end{equation}
This implies that $\ell(\mathbf{a},\mathbf{b})\geq s$.

Next, we discuss the solutions to the linear system (\ref{fdsec}).

{\bf Case 1:} $n\leq 2k$. We show that there exist two distinct codewords $\mathbf{a},\mathbf{b}\in \C$ such that $\ell(\mathbf{a},\mathbf{b})=n-1$. Let $I,J$ be any two increasing vectors of $[n]^{n-1}$, that is, $s=n-1$ in the matrix given in (\ref{fdfir}). The rank of matrix $M_{IJ}$ defined in (\ref{fdfir}) is less than $2k$. So, the corresponding linear system (\ref{fdsec}) has nonzero solutions. Moreover, there exist two increasing vectors $I,J\in [n]^{n-1}$ such that the solution $(\mathbf{x},-\mathbf{y})$ of the corresponding linear system (\ref{fdsec}) satisfies $\mathbf{x}\neq \mathbf{y}$. In this case, there exist two distinct codewords $\mathbf{a}=\mathbf{x}G$ and  $\mathbf{b}=\mathbf{y}G$ satisfying $\ell(\mathbf{a},\mathbf{b})=n-1$ since $\mathbf{a}\neq \mathbf{b}$. Otherwise, we choose $I=(1,2,\cdots,n-1)$ and $J=(2,3,\cdots,n)$, and let $(\mathbf{x},-\mathbf{y})\in \mathbb{F}_q^{2k}$ be a nonzero solution of the linear system (\ref{fdsec}) with $\mathbf{x}=\mathbf{y}$. This gives
\begin{equation}\notag
 (f_1(\mathbf{x}),f_2(\mathbf{x}),\cdots,f_{n-1}(\mathbf{x}))=(f_2(\mathbf{y}),f_3(\mathbf{y}),\cdots,f_n(\mathbf{y}))=(f_2(\mathbf{x}),f_3(\mathbf{x}),\cdots,f_n(\mathbf{x})).
\end{equation}
So, $f_1(\mathbf{x})=f_2(\mathbf{x})=\cdots=f_{n}(\mathbf{x})$, i.e., $(1,1,\cdots,1)\in \C$. This is a contradiction.

{\bf Case 2:} $n>2k$. We show that there exist two distinct codewords $\mathbf{a},\mathbf{b}\in \C$ such that $\ell(\mathbf{a},\mathbf{b})\geq 2k-1$. Let $I$, $J$ be any two increasing vectors of $[n]^{2k-1}$, that is, $s=2k-1$ in the matrix given in (\ref{fdfir}). The rank of $M_{IJ}$ given in (\ref{fdfir}) is less than $2k$, and so the corresponding linear system (\ref{fdsec}) has nonzero solutions. Moreover, there exist two increasing vectors $I,J\in [n]^{2k-1}$ such that the corresponding linear system (\ref{fdsec}) has a nonzero solution $(\mathbf{x},-\mathbf{y})\in \mathbb{F}_q^{2k}$ satisfying $\mathbf{x}\neq \mathbf{y}$. In this case, the code $\C$ has two distinct codewords $\mathbf{a}=\mathbf{x}G$ and $\mathbf{b}=\mathbf{y}G$ in $\C$ satisfying $\ell(\mathbf{a},\mathbf{b})\geq 2k-1$. Otherwise, assume that for any two increasing vectors $I,J\in [n]^{2k-1}$, the corresponding linear system (\ref{fdsec}) has only solutions with the form $(\mathbf{x},-\mathbf{x})\in \mathbb{F}_q^{2k}$. Then we will derive a contradiction.

 Choose $I=(1,2,\cdots,2k-1)$ and $J=(2,3,\cdots,2k)$. If the corresponding linear system (\ref{fdsec}) has only nonzero solutions of the form $(\mathbf{x},-\mathbf{x})\in \mathbb{F}_q^{2k}$, then we have
 \begin{equation}\notag
 (f_1(\mathbf{x}),f_2(\mathbf{x}),\cdots,f_{n-1}(\mathbf{x}))=(f_2(\mathbf{y}),f_3(\mathbf{y}),\cdots,f_n(\mathbf{y}))=(f_2(\mathbf{x}),f_3(\mathbf{x}),\cdots,f_n(\mathbf{x})).
 \end{equation}
 So, $f_1(\mathbf{x})=f_2(\mathbf{x})=\cdots=f_{2k}(\mathbf{x})$. This further shows that the code $\C$ has a codeword of the form
 \begin{equation}\notag
\mathbf{c}^{(1)}=(\underbrace{1,1,\cdots,1}_{2k},\ast,\ast,\cdots,\ast).
\end{equation}
Similarly, choosing $I=(2,3,\cdots,2k)$ and $J=(3,4,\cdots,2k+1)$, we can derive that the code $\C$ has a codeword of the form
\begin{equation}\notag
\mathbf{c}^{(2)}=(\star,\underbrace{1,1,\cdots,1}_{2k},\star,\star,\cdots,\star).
\end{equation}
Repeating the above process $n-2k+1$ times, we get the $(n-2k+1)$th codeword in $\C$ of the form
\begin{equation}\notag
\mathbf{c}^{(n-2k+1)}=(\diamond,\diamond,\cdots,\diamond,\underbrace{1,1,\cdots,1}_{2k}).
\end{equation}
Suppose that $\mathbf{u},\mathbf{v}\in \mathbb{F}_q^k$ are the message vectors of $\mathbf{c}^{(1)}$ and $\mathbf{c}^{(2)}$, respectively. Then we have
\begin{equation}\notag
(\mathbf{u}G_2,\mathbf{u}G_3,\cdots,\mathbf{u}G_{2k})=(\mathbf{v}G_2,\mathbf{v}G_3,\cdots,\mathbf{v}G_{2k})=(1,1,\cdots,1),
\end{equation}
 where $G_i's$ are columns of the generator matrix of $G$. This implies that
 \begin{equation}\label{fdthr}
 (\mathbf{u}-\mathbf{v})(\underbrace{G_2,G_3,\cdots,G_{2k}}_{\bar{G}})=(0,0,\cdots,0).
 \end{equation}
 Since $n-d<2k-1$, i.e., $n-d+1\leq 2k-1$, by Lemma \ref{lemtr} there exist $k$ columns of $\bar{G}$ are linearly independent over $\mathbb{F}_q$, i.e., the rank of $\bar{G}$ is equal to $k$. So, the linear system (\ref{fdthr}) has only zero solution, i.e., $\mathbf{u}=\mathbf{v}$. This leads to $\mathbf{c}^{(1)}=\mathbf{c}^{(2)}$. Repeating above discussion we derive that the code $\C$ has a codeword
 \begin{equation}\notag
 \mathbf{c}^{(1)}=\mathbf{c}^{(2)}=\cdots=\mathbf{c}^{(n-2k+1)}=(1,1,\cdots,1).
 \end{equation}
 This is a contradiction.         \EOP

Theorem \ref{thm:upperbound} and Lemma~\ref{lemChenbound} show that the insdel distance of a linear code will be affected by whether it contains the codeword $\mathbf{1}$.
This fact can be verified by the following simple example. Let $\C$ be an $[n, 1]$ code over $\bF_q$, then $\C = \{ a \mathbf{c} \,:\, a\in \bF_q \}$, where $\mathbf{c}\in \bF_q^n$. If $\mathbf{1} \in \C$, then $\C=\{ a\cdot \mathbf{1}\,:\, a \in \bF_q\}$. For any two distinct codewords $\mathbf{c}_1, \mathbf{c}_2 \in \C$, $\ell(\mathbf{c}_1,\mathbf{c}_2)=0$, and so,
$d_I(\C)=2n$. In this case, the insdel distance of the code $\C$ reaches the half-Singleton bound given in Lemma \ref{lemChenbound}. If $\mathbf{1}\notin \C$, then we can show that there always exist two codewords $\mathbf{c}_1, \mathbf{c}_2 \in \C$ such that $\ell(\mathbf{c}_1,\mathbf{c}_2)\geq 1$, and so, $d_I(\C)\leq 2(n-1)$. In this case, the insdel distance of the code $\C$ may reach the upper bound given in Theorem \ref{thm:upperbound}, but never reach the upper bound given in Lemma~\ref{lemChenbound}.

By Theorem~\ref{thm:upperbound}, we give another proof of the half-Singleton bound on the insdel distance of linear codes.

\begin{corollary}\label{corhalfbound}
Let $\C$ be an $[n,k]$ linear code over $\bF_q$. Its insdel distance satisfies
\begin{equation}\notag
d_I(\C) \leq \max\left\{  2(n-2k+2),\, 2 \right\}.
\end{equation}
\end{corollary}

{\it Proof:} If $\mathbf{1}\notin\C$, by Theorem~\ref{thm:upperbound}, we have $d_I(\C) \leq \max\left\{  2(n-2k+1),\, 2 \right\}\leq \max\left\{  2(n-2k+2),\, 2 \right\}$. If $\mathbf{1}\in \C$, we only need to prove that there exist two distinct codewords $\mathbf{a},\mathbf{b}\in \C$ satisfying $\ell(\mathbf{a},\mathbf{b})\geq \min\left\{  2k-2,\, n-1 \right\}$. Since $\mathbf{1}\in \C$,  $\C$ has a generator matrix as the following form:
\begin{equation}\label{matrixGn-1}
G=\begin{pmatrix}
\mathbf{1}_{n-1} & 1\\
 G_{n-1}  & \mathbf{0}_{n-1}^T
\end{pmatrix},
\end{equation}
where $G_{n-1}$ is a $(k-1)\times(n-1)$ matrix over $\mathbb{F}_q$, $\mathbf{1}_{n-1}=(\underbrace{1,1,\cdots,1}_{n-1})$ and $\mathbf{0}_{n-1}=(\underbrace{0,0,\cdots,0}_{n-1})$. Let $\C_{n-1}$ be the $[n-1,k-1]$ linear code generated by $G_{n-1}$. If $\mathbf{1}_{n-1}\in \C_{n-1}$, then $\mathbf{1}^{\prime}=(\mathbf{1}_{n-1},0)\in \C$ and $\ell(\mathbf{1}^{\prime},\mathbf{1})=n-1\geq \min\left\{ 2k-2,\, n-1 \right\}$. If $\mathbf{1}_{n-1}\notin \C_{n-1}$, by Theorem~\ref{thm:upperbound}, there exist two distinct codewords $\mathbf{a}_{n-1},\mathbf{b}_{n-1}\in \C_{n-1}$ such that $\ell(\mathbf{a}_{n-1},\mathbf{b}_{n-1})\geq \min\left\{  2k-3,\, n-2 \right\}$. It is clear that $\mathbf{a}=(\mathbf{a}_{n-1},0)\in \C$, $\mathbf{b}=(\mathbf{b}_{n-1},0)\in \C$, and $\ell(\mathbf{a},\mathbf{b})\geq \min\left\{  2k-2,\, n-1 \right\}$.
\EOP

For a linear code $\C$ over $\bF_q$, we know that $d_I(\C)=2$ if $n<2k$ by Lemma~\ref{lemChenbound}. These codes can not correct insdel errors. When $n=2k$, from Lemma~4 in \cite{Abdel2010} we know that the following linear
insdel code attains the half-Singleton bound.
\begin{corollary}\label{cor1}
For a positive integer $k$, let $\C$ be an $[2k,k]$ code over $\mathbb{F}_q$ given by
\[ \C=\left\{ (c_1,c_2,\cdots,c_{2k}): c_i=c_{2k-i+1}\in \mathbb{F}_q, \,\, i=1,2,\cdots, k \right\}. \]
Then $d_I(\C) =4$, i.e., $\C$ is optimal with respect to the half-Singleton bound.
\end{corollary}

\begin{remark}
The length of the optimal linear insdel codes given in Corollary~\ref{cor1} is independent of the size of the finite field.
\end{remark}

The following corollary shows that only in very special cases, a linear code $\C$ and its dual $\C^\perp$ have the insdel error-correcting capability at the same time.
This result directly follows from Theorem \ref{thm:upperbound} and Lemma~\ref{lemChenbound}.

\begin{corollary}\label{cor2}
Let $\C$ be an $[n, k]$ code over $\mathbb{F}_q$ and $\C^{\perp}$ be its dual code. If both $\C$ and $\C^\perp$ have insdel error-correcting capability, then $n=2k$.
In this case, $\mathbf{1}\in \C$, $d_I(\C) = d_I(\C^\perp) =4$ and $p \, |\, n$, where $p$ is the characteristic of the field $\bF_q$.
\end{corollary}

%
%
%

\section{Optimal linear insdel codes attaining the strict half-Singleton bound}

In this section, we present a sufficient condition for a linear insdel code to be optimal according to the strict half-Singleton bound given in Theorem~\ref{thm:upperbound}. Then we give several examples of optimal linear insdel codes. To this end, we first introduce some useful notation. Let $n, k$ be positive integers with $2k< n$. Let $I, J\in [n]^{2k}$ be increasing vectors with length $2k$. Let $I\cap J$ be a increasing vector made up of the corresponding equal components of $I$ and $J$, i.e., $I\cap J = (r_1, r_2, \cdots, r_t)$, $t\leq 2k$, where $r_i= I_{e_i} = J_{e_i}$, which is the $e_i$th component of $I$ and $J$ for $1\leq i\leq t$.

\begin{theorem}\label{theorem2}
Let $\C$ be an $[n,k]$ code over $\bF_q$ with generator matrix $G=(G_1, G_2, \cdots, G_n)$, where $G_i$ is the $i$th column of $G$ and $n>2k$. If for every two increasing vectors $I, J\in [n]^{2k}$ with $rank(G_{I\cap J})<k$,
where $G_{I\cap J} = (G_{e_1}, G_{e_2}, \cdots, G_{e_t})$ and $I\cap J =(e_1, e_2 , \cdots, e_t)$,
it holds that $det(M_{IJ})\neq 0$, where
\begin{equation}\notag
M_{IJ}=
\begin{pmatrix}
G_{I_1} & G_{I_2} & \cdots & G_{I_{2k}}\\
G_{J_1} & G_{J_2} & \cdots & G_{J_{2k}}
\end{pmatrix},
\end{equation}
then $d_I(\C) = 2(n-2k+1)$, i.e., $\C$ is optimal with respect to the strict half-Singleton bound.
\end{theorem}

{\it Proof:} First, we show that $\mathbf{1}\notin \C$. Otherwise, assume that $\mathbf{1}\in \C$, then it has a generator matrix as the following form:
\begin{equation}\notag
G^\prime=
\begin{pmatrix}
1 & 1 & \cdots & 1\\
g_{21}^\prime & g_{22}^\prime & \cdots & g_{2n}^\prime\\
\vdots & \vdots & \quad & \vdots\\
g_{k1}^\prime & g_{k2}^\prime & \cdots & g_{kn}^\prime
\end{pmatrix}=\begin{pmatrix}G_1^\prime & G_2^\prime & \cdots & G_n^\prime\end{pmatrix}.
\end{equation}
Consider the matrix
\begin{equation}\notag
M_{IJ}^\prime=\begin{pmatrix}G_{I_1}^\prime & G_{I_2}^\prime & \cdots & G_{I_{2k}}^\prime\\G_{J_1}^\prime & G_{J_2}^\prime & \cdots & G_{J_{2k}}^\prime\end{pmatrix}
\end{equation}
for two increasing vectors $I,J\in [n]^{2k}$ with $rank(G_{I\cap J})<k$. Since $M_{IJ}^\prime$ has two rows with all $1$s, $\det(M_{IJ}^\prime)=0$. On the other hand, there exists a $k\times k$ invertible matrix $Q$ such that $G=QG^\prime$. Let
\begin{equation}\notag
N=\begin{pmatrix}Q & 0_{k\times k}\\0_{k\times k} & Q\end{pmatrix}.
\end{equation}
It is easy to verify that $NM_{IJ}^\prime=M_{IJ}$. Then $\det(M_{IJ})=\det(N) \det(M_{IJ}^\prime)=0$. This is a contradiction. It follows that $\mathbf{1}\notin \C$.
By Theorem~\ref{thm:upperbound}, $d_I(\C) \leq 2(n-2k+1)$.

Second, we show that for any two different codewords $\mathbf{a}, \mathbf{b}\in \C$, $\ell(\mathbf{a}, \mathbf{b})\leq 2k-1$. Otherwise, assume that there exist two distinct codewords $\mathbf{a}, \mathbf{b}$ such that $\ell(\mathbf{a}, \mathbf{b})\geq 2k$, then there exist two increasing vectors $I,J\in [n]^{2k}$ such that $\mathbf{a}_I = \mathbf{b}_J$, i.e., $\mathbf{a}_{I_s} = \mathbf{b}_{J_s}$ for $s=1, 2, \cdots, 2k$. Let $\mathbf{x}, \mathbf{y}\in \bF_q^k$ be the message symbols of the codewords $\mathbf{a}, \mathbf{b}$ respectively, i.e., $\mathbf{a}= \mathbf{x}G$ and $\mathbf{b} = \mathbf{y} G$.
From assumption we see that the linear system
\[ \mathbf{z} G_{I\cap J} = (\mathbf{a}_{e_1}, \mathbf{a}_{e_2}, \cdots, \mathbf{a}_{e_t})\]
has two distinct solutions $\mathbf{x}$ and $\mathbf{y}$. So, ${\rm rank}(G_{I\cap J})<k$. Since $\mathbf{a}_I = \mathbf{b}_J$, we have
\begin{equation}\notag
(\mathbf{x}, -\mathbf{y} )M_{IJ}=0.
\end{equation}
So, $\det(M_{IJ})=0$. This contradicts the assumption in the theorem. Thus, it follows that $d_I(\C) \geq 2(n-2k+1)$, and then $d_I(\C)=2(n-2k+1)$ by Theorem~\ref{thm:upperbound}.  \EOP

Next we use Theorem~\ref{theorem2} to give some examples of optimal linear insdel codes.

\begin{example}\label{ex1}
Let $q=49$ and $w$ be a generator of $\bF_q$. Let $\C$ be an $[5, 2]$ code over $\bF_q$ with generator matrix
\[G=\begin{pmatrix}w^{28} & w & w^{39} & w^{26} & w^{20}\\
w^{10} & w^{13} & 2 & w^{37} & w
\end{pmatrix}.\]
It is easy to see that any two columns of $G$ are linear independent over $\bF_q$. Two different increasing vectors $I, J\in [5]^4$ satisfying ${\rm rank}(G_{I\cap J})<2$
if and only if $I\cap J=(\emptyset), (1)$ or $(5)$. So, all possible cases of the vectors $I$ and $J$ are as follows: $I=(1,2,3,4)$ and $J=(2,3,4,5)$; $I=(1,2,3,4)$ and $J=(1,3,4,5)$;
$I=(1,2,3,5)$ and $J=(2,3,4,5)$.  The corresponding matrices $M_{IJ}$ are as follows:
\begin{equation}
M_{IJ}=\begin{pmatrix}\notag
w^{28}  &  w      &   w^{39}  &  w^{26}\\
w^{10}  & w^{13}  &    2      &  w^{37}\\
w       & w^{39}  &   w^{26}  &  w^{20}\\
w^{13}  &  2      &   w^{37}  &   w    \\
\end{pmatrix},
\begin{pmatrix}
w^{28}  &  w      &   w^{39}  &  w^{26}\\
w^{10}  & w^{13}  &    2      &  w^{37}\\
w^{28}  & w^{39}  &   w^{26}  &  w^{20}\\
w^{10}  &  2      &   w^{37}  &   w    \\
\end{pmatrix},
\begin{pmatrix}
w^{28}  &  w      &   w^{39}  &    w^{20}\\
w^{10}  & w^{13}  &    2      &      w   \\
w       & w^{39}  &   w^{26}  &    w^{20}\\
w^{13}  &  2      &   w^{37}  &      w   \\
\end{pmatrix}.
\end{equation}
By help of Magma one easily show that $\det (M_{IJ})\neq 0$ for above three cases. From Theorem~\ref{theorem2} we know that $d_I(\C)=4$.
In fact, we find two codewords $\mathbf{a}=(w^{38},w^{14},w^7,w^{15},w^{21}), \mathbf{b}=(w^2,w^{38},w^{14},w^7,w^2)$ satisfying $\ell(\mathbf{a}, \mathbf{b})=3$, and so
$d_I(\mathbf{a}, \mathbf{b})=4$.
\end{example}

\begin{example}\label{ex3}
Let $q=121$, and $w$ be a generator of $\mathbb{F}_q$. Let $\C$ be an $[8,3]$ code over $\bF_q$ with generator matrix
\begin{equation}\notag
G=\begin{pmatrix}
w^{40} &  w^{20}  &  w^{22}  &   w^3      &  w^{49}  &  w^{55} & w^{54} & w^{65}\\
w^{86} &  w^{27}  &  w^{89}  &   w^{64}   &  w^{73}  &  w^{23} & w^{44} & w^{79}\\
w^{88} &  w^{103} &  w^{110} &   w^{97}   &  w^{21}  &  w^{51} & w^{47} & w^{70}
\end{pmatrix}.
\end{equation}
By help of Magma, we can verify that $\det(M_{IJ}) \neq 0$ for all two different increasing vectors $I, J\in [8]^4$ that satisfy ${\rm rank}(G_{I\cap J})<3$.  From Theorem~\ref{theorem2}, we know that $d_I(\C)=6$.
In fact, we find two codewords $\mathbf{a}=(w^{95},w,w^{2},w^{80},w^{67},w^{40},w^{31},w^{79}), \mathbf{b}=(6,w^{95},w,w^{2},w^{80},w^{67},w^6,w^{112})$ satisfying $\ell(\mathbf{a}, \mathbf{b})=5$, and so
$d_I(\mathbf{a}, \mathbf{b})=6$.
\end{example}

\begin{example}\label{ex4}
Let $q=169$, and $w$ be a generator of $\mathbb{F}_q$. Let $\C$ be an $[9, 4]$ code over $\bF_q$ with generator matrix
\begin{equation}\notag
G=\begin{pmatrix}
w^{81} & w^{120} &  w^4   & w^{136}  &  w^{147} & w^{71} & w^{166} & w^{132} & w^{103}\\
w^{83} & w^{155} & w^{82} & w^{163}  &  w^{48}  & w^{36} & w^{88}  & w^{63}  & w^{45}\\
w^{143} &  w^{85} & w^{72}  &  w^{146} & w^{117} & w^{18} & w^{95}  & w^{12} & w^{134}\\
w^{131} & w^{160} & w^{27}  &  w^{148} & w^{164} &  w^7   & w^{109} & w^{107} & w^{32}
\end{pmatrix}.
\end{equation}
By help of Magma, we can verify that $\det(M_{IJ}) \neq 0$ for all two different increasing vectors $I, J\in [9]^4$ that satisfy ${\rm rank}(G_{I\cap J})<4$.
From Theorem \ref{theorem2}, we know $d_I(\C)=4$. In fact, we find two codewords
$\mathbf{a}=(w^{9},w^{127},w^{13},w^{22},w^{21},w^{11},w^{53},w^{165},w^{110}),$ $\mathbf{b}=(w^{120},w^{9},w^{127},w^{13},w^{22},w^{21},w^{11},w^{53},7)$
satisfying $\ell(\mathbf{a}, \mathbf{b})=7$, and so $d_I(\mathbf{a}, \mathbf{b})=4$.
\end{example}

In the following, we present a class of optimal $[2k+1, k]$ linear insdel codes over $\bF_q$ for some positive integer~$k$. To this end, we first introduce some notation.
Let $t$ be a positive integer with $t\leq k$ and $\Omega_t = \{ t, t+1, \cdots, k\}$. Denote by $\Omega_t^o= \{ i \in \Omega_t \,|\, k-i \,\, {\rm is \,\, odd}\,\, \}$,
$\Omega_t^e= \{ i \in \Omega_t \,|\, k-i \,\, {\rm is \,\, even}\,\,\}$ and
\begin{equation}\label{eq:generator}
G=\begin{pmatrix}
1 & 0 & \cdots & 0 & a_1 & 0 & \cdots & 0 & 1\\
0 & 1 & \cdots & 0 & a_2 & 0 & \cdots & 1 & 0\\
\vdots & \vdots & \quad & \vdots & \vdots & \vdots & \quad & \vdots & \vdots \\
0 & 0 & \cdots & 1 & a_k & 1 & \cdots & 0 & 0\\
\end{pmatrix}_{k\times (2k+1)},
\end{equation}
where $a_i\in \bF_q$ for $i=1, 2, \cdots, k$ satisfy $\sum_{i=1}^k a_i \neq 1$.

\begin{proposition}\label{pro}
Let symbols be given as above and $\C$ be an $[2k+1, k]$ code with generator matrix $G$ given in (\ref{eq:generator}). If for any $t$ with $1\leq t\leq k$ satisfies
$$\sum_{i\in \Omega_t^o} a_i-\sum_{i\in \Omega_t^e} a_i \neq 1,$$
then $d_I(\C) =4$, i.e., $\C$ is optimal with respect to the strict half-Singleton bound.
\end{proposition}

{\it Proof:} Since $\sum_{i=1}^ka_i\neq 1$, it is easy to verify that $ \mathbf{1} \notin \C$. So, $d_I(\C)\leq 2(2k+1-2k+1)=4$ by Theorem \ref{thm:upperbound}.
In the following, we show that $d_I(\C)\neq 2$.

Assume $d_I(\C)=2$, then the linear code $\C$ contains a codeword $\mathbf{c}=(c_1,c_2,\cdots,c_n)$ and a nonzero codeword $\mathbf{x}=(x_1,x_2,\cdots,x_n)$ as characterized in Lemma \ref{lemsec} for some
$u$, $v$ and $\alpha$, where $n=2k+1$. Since $\mathbf{x}$ is nonzero, from the representation of codewords in $\C$ we know that $x_i\neq 0$ for some $i< k+1$. Thus $u<k+1$ and $v > k+1$. Let $t=\min\{i\,|\, i\in [n]\,\, {\rm and} \,\, x_i\neq 0\}$ and $t^\prime=\max\{i\,|\, i\in [n]\,\,{\rm  and} \,\, x_i\neq 0\}$, then $t^\prime=n-t+1$, $1\leq t\leq k$ and $u\leq t< k+1 <t^\prime\leq v$. Thus,
$$ x_i=\left\{
\begin{array}{lrl}
0,          &  & for\ 1\leq i<t \ or\ t^{\prime}<i\leq n\\
c_{i+1}-c_i,&  & for\ t\leq i<t^{\prime}\\
\beta,          &  & for\ i=t^{\prime}
\end{array},
\right.
$$
where $\beta\in \bF_q$. From the generator matrix of $\C$ we know that the codeword $\mathbf{c}=(c_1,c_2,\cdots,c_n)\in\C$ satisfies that
\begin{equation}\label{eq:codeform}
 c_{k+1} = \sum_{i=1}^k a_i c_i, \,\, c_j = c_{n+1-j}, \,\, j= k+2, k+3, \cdots, n.
\end{equation}
Then the codeword $\mathbf{y}=\mathbf{c} +\mathbf{x}=(y_1, y_2, \cdots, y_n)\in \C$ satisfies that
$$ y_i=\left\{
\begin{array}{lrl}
c_i,          &  & for\ 1\leq i<t \ or\ t^{\prime}<i\leq n\\
c_{i+1},&  & for\ t\leq i<t^{\prime}\\
c_t+\beta,          &  & for\ i=t^{\prime}
\end{array}.
\right.
$$
Since any codeword in $\C$ satisfies the relation given in (\ref{eq:codeform}), from the representation of $\mathbf{y}$ we have that
$$c_t=c_{t+2}=\cdots=c_{k-1}=c_{k+1}\,\, {\rm and} \,\, c_{t+1}=c_{t+3}=\cdots=c_{k-2}=c_{k}$$
if $k-t$ is odd, and
$$c_t=c_{t+2}=\cdots=c_{k-2}=c_{k}\,\, {\rm and}\,\, c_{t+1}=c_{t+3}=\cdots=c_{k-1}=c_{k+1}$$
if $k-t$ is even. So, for the codeword $\mathbf{c}$ we have
\begin{equation}\label{eq:codewordc}
c_{k+1}=\sum_{i=1}^{t-1}c_ia_i+c_{k}\sum_{i\in \Omega_t^e} a_i+c_{k+1}\sum_{i\in \Omega_t^o} a_i,
\end{equation}
and for the codeword $\mathbf{y}$ we have
\begin{equation}\label{eq:codewordy}
c_k=\sum_{i=1}^{t-1}c_ia_i+c_{k}\sum_{i\in \Omega_t^o}a_i+c_{k+1}\sum_{i\in \Omega_t^e}a_i.
\end{equation}
By $(\ref{eq:codewordc})-(\ref{eq:codewordy})$, we have
\begin{equation}\label{eq:ckck1}
 c_{k+1}-c_{k}=(c_{k+1}-c_{k})\left( \sum_{i\in \Omega_t^o}a_i-\sum_{i\in \Omega_t^e}a_i \right).
\end{equation}
Since $\mathbf{x}$ is a nonzero codeword, we know that $c_{k+1} \neq c_{k}$. From (\ref{eq:ckck1}) we have
\[  \sum_{i\in \Omega_t^o}a_i-\sum_{i\in \Omega_t^e}a_i  =1. \]
This is a contradiction. So, $d_I(\C) =4$, and $\C$ is an optimal linear insdel code with respect to the strict half-Singleton bound.

\begin{remark}\label{remarkcon}
When $q>2$, one can verify that for any $t$ with $1\leq t\leq k$, there exist $a_i\in \bF_q$ satisfying
\begin{equation}\label{eq:ai}
 \sum_{i=1}^k a_i \neq 1 \,\, {\rm and} \,\, \sum_{i\in \Omega_t^o}a_i-\sum_{i\in \Omega_t^e}a_i  \neq 1 .
\end{equation}
For example, $a_{k-1}\in \bF_q\setminus \{0, 1\}$ and $a_i=0$ for all $i\in \{ 1, 2, \cdots, k\}\setminus \{ k-1\}$
satisfy (\ref{eq:ai}). So, there exist optimal $[2k+1, k, 4]$ linear insdel codes over $\bF_q$ if $q>2$. Moreover,
the length of $\C$ is independent of the size of the finite field $\bF_q$.
\end{remark}

\section{Optimal binary linear insdel codes attaining the (strict) half-Singleton bound}

In this section we study optimal linear insdel codes over $\mathbb{F}_2$ with respect to the half-Singleton bound and the strict half-Singleton bound proposed in Theorem~\ref{thm:upperbound}, respectively.

\begin{lemma}\label{2n3}
For a positive integer $k$, let $\C$ be an $[2k+3,k]$ linear insdel code over $\mathbb{F}_2$ without codeword with $2k$ consecutive coordinates being 1. Then there exist two distinct codewords $\mathbf{u}, \mathbf{v}\in \C$ such that $\ell(\mathbf{u}, \mathbf{v})\geq 2k$.
\end{lemma}

{\it Proof:} Let $d_H$ be the minimal Hamming distance of $\C$, then there exist a codeword $\mathbf{z}\in \C$ such that $w_H(\mathbf{z})=d_H$. So $\ell(\mathbf{z},\mathbf{0})=2k+3-d_H$. If $d_H\leq 3$, then the conclusion follows. Next we discuss the case of $d_H>3$.

Let $G=(G_1, G_2, \cdots, G_{2k+3})$ be a generator matrix of $\C$. 
Consider the linear equations
\begin{equation}\label{eq:linsys1}
(\mathbf{x}, -\mathbf{y})\underbrace{\begin{pmatrix}G_{1} & G_{2} & \cdots & G_{2k-1}\\
 G_{2} & G_{3} & \cdots & G_{2k}\end{pmatrix}}_M =0.
\end{equation}
The rank of $M$ is less than $2k$. So, the linear system (\ref{eq:linsys1}) has a nonzero solution $(\mathbf{x_1},-\mathbf{y_1})\in \mathbb{F}_2^{2k}$. Moreover, we claim that $\mathbf{x_1}\neq \mathbf{y_1}$. Otherwise,
if $\mathbf{x_1}= \mathbf{y_1}${\blue,} then from (\ref{eq:linsys1}) we have
\[ \left( f_1(\mathbf{x_1}),f_2(\mathbf{x_1}),\cdots,f_{2k-1}(\mathbf{x_1})\right)=\left( f_2(\mathbf{y_1}),f_3(\mathbf{y_1}),\cdots,f_{2k}(\mathbf{y_1})\right)
=\left( f_2(\mathbf{x_1}),f_3(\mathbf{x_1}),\cdots,f_{2k}(\mathbf{x_1})\right),\]
where $f_i(\mathbf{x})= \mathbf{x}G_i$ for $1\leq i\leq 2k$. So, $f_1(\mathbf{x_1})=f_2(\mathbf{x_1})=\cdots=f_{2k}(\mathbf{x_1})$. Since $d_H(\C)>3$, we derive that $\C$
has a codeword of the form $(\underbrace{1,1,\cdots,1}_{2k},\ast,\ast,\ast)$. This is a contradiction.

Let $\mathbf{x_1}, \mathbf{y_1}$ be message symbols of codewords $\mathbf{a}, \bar{\mathbf{a}}$, respectively. Then they are different and have the following form:
$$\mathbf{a}=\left(a_1,a_2,a_3,a_4,\cdots,a_{2k-1},\alpha_1,\alpha_2,\alpha_3,\alpha_4\right), \,\, \bar{\mathbf{a}}=\left(\bar{\alpha}_1,a_1,a_2,a_3,\cdots,a_{2k-2},a_{2k-1},\bar{\alpha}_2,\bar{\alpha}_3,\bar{\alpha}_4\right).$$
It is clear that the length of the longest common subsequence of $\mathbf{a}$ and $\bar{\mathbf{a}}$ is at least $2k-1$. If there exist some $i\in \{1,2,3,4\}$ and $j\in \{2,3,4\}$ such that $\alpha_i=\bar{\alpha}_j$, then $\ell(\mathbf{a},\bar{\mathbf{a}})\geq 2k$, and the conclusion follows. If $\alpha_i\neq \bar{\alpha}_j$ for any $i\in \{1,2,3,4\}$ and $j\in \{2,3,4\}$, we have
\begin{equation}\label{alpha1}
\alpha_i+\bar{\alpha}_j=1.
\end{equation}

By choosing proper matrix $M$ as in (\ref{eq:linsys1}) we can show that $\C$ has two different codewords $\mathbf{b}$ and $\bar{\mathbf{b}}$ as the following form:
$$\mathbf{b}=(\beta_1,b_1,b_2,b_3,\cdots,b_{2k-2},b_{2k-1},\beta_2,\beta_3,\beta_4),\,\, \bar{\mathbf{b}}=(\bar{\beta}_1,\bar{\beta}_2,b_1,b_2,\cdots,b_{2k-3},b_{2k-2},b_{2k-1},\bar{\beta}_3,\bar{\beta}_4).$$
It is clear that the length of the longest common subsequence of $\mathbf{b}$ and $\bar{\mathbf{b}}$ is at least $2k-1$. If $\beta_1\in \{ \bar{\beta}_1,\bar{\beta}_2\}$ or $\{ \beta_2,\beta_3,\beta_4\} \cap
\{ \bar{\beta}_3,\bar{\beta}_4\}\neq \emptyset$, then $\ell(\mathbf{b},\bar{\mathbf{b}})\geq 2k$, and the conclusion follows. Otherwise, for $j\in \{1, 2\}$, $u\in \{2, 3, 4\}$ and $v\in \{3, 4\}$
we have
\begin{equation}\label{beta1}
\beta_1 + \bar{\beta_j} = 1 \,\, {\rm and }\,\, \beta_u + \bar{\beta}_v=1.
\end{equation}
Since $\C$ is a linear code, $\mathbf{a}+\mathbf{b}$ and $\bar{\mathbf{a}}+\bar{\mathbf{b}}$ are also codewords of $\C$ and have the following form:
\begin{equation}\notag
\mathbf{a}+\mathbf{b}=(a_1+\beta_1,a_2+b_1,a_3+b_2,a_4+b_3,\cdots,a_{2k-1}+b_{2k-2},\alpha_1+b_{2k-1},\alpha_2+\beta_2,\alpha_3+\beta_3,\alpha_4+\beta_4),
\end{equation}
\begin{equation}\notag
\bar{\mathbf{a}}+\bar{\mathbf{b}}=(\bar{\alpha}_1+\bar{\beta}_1,a_1+\bar{\beta}_2,a_2+b_1,a_3+b_2,\cdots,a_{2k-2}+b_{2k-3},a_{2k-1}+b_{2k-2},\bar{\alpha}_2+b_{2k-1},\bar{\alpha}_3+\bar{\beta}_3,\bar{\alpha}_4+\bar{\beta}_4).
\end{equation}
If $\mathbf{a}+\mathbf{b}\neq \bar{\mathbf{a}}+\bar{\mathbf{b}}$, by (\ref{alpha1}) and (\ref{beta1}) we have that $\ell(\mathbf{a}+\mathbf{b}, \bar{\mathbf{a}}+\bar{\mathbf{b}})\geq 2k$, and the conclusion follows.
If $\mathbf{a}+\mathbf{b}=\bar{\mathbf{a}}+\bar{\mathbf{b}}$, by (\ref{alpha1}) and (\ref{beta1}), then we can derive that $\mathbf{a}+\mathbf{b}=(0,\underbrace{1,1,\cdots,1}_{2k-1},0,0,0)$ or $(1,\underbrace{0,0,\cdots,0}_{2k-1},1,1,1)$.

Again by similar analysis of the beginning of the proof, we can show that $\C$ has two different codewords $\mathbf{c}$ and $\bar{\mathbf{c}}$ as the following form:
$$\mathbf{c}=(\gamma_1,\gamma_2,c_1,c_2,c_3,\cdots,c_{2k-2},c_{2k-1},\gamma_3,\gamma_4),\,\, \bar{\mathbf{c}}=(\bar{\gamma}_1,\bar{\gamma}_2,\bar{\gamma}_3,c_1,c_2,\cdots,c_{2k-3},c_{2k-2},c_{2k-1},\bar{\gamma}_4).$$
It is clear that the length of the longset common subsequence of $\mathbf{c}$ and $\bar{\mathbf{c}}$ is at least $2k-1$. If $\{ \gamma_1,\gamma_2 \} \cap \{\bar{\gamma}_1,\bar{\gamma}_2,\bar{\gamma}_3\} \neq \emptyset$ or
$\bar{\gamma}_4 \in \{ \gamma_3,\gamma_4 \}$, then $\ell(\mathbf{c},\bar{\mathbf{c}})\geq 2k$, and the conclusion follows. Otherwise, for $i\in \{ 3, 4\}$, $s\in \{1, 2\}$ and $t\in \{1,2,3\}$ we have
\begin{equation}\label{gamma1}
\gamma_i+\bar{\gamma}_4=1 \,\, {\rm and }\,\, \gamma_s +\bar{\gamma}_t =1.
\end{equation}
Since $\C$ is a linear code, $\mathbf{b}+\mathbf{c}$ and $\bar{\mathbf{b}}+\bar{\mathbf{c}}$ are also codewords of $\C$ and have the following form:
\begin{equation}\notag
\mathbf{b}+\mathbf{c}=(\beta_1+\gamma_1,b_1+\gamma_2,b_2+c_1,b_3+c_2,\cdots,b_{2k-1}+c_{2k-2},\beta_2+c_{2k-1}, \beta_3+\gamma_3,\beta_4+\gamma_4)
\end{equation}
and
\begin{equation}\notag
\bar{\mathbf{b}}+\bar{\mathbf{c}}=(\bar{\beta}_1+\bar{\gamma}_1,\bar{\beta}_2+\bar{\gamma}_2,b_1+\bar{\gamma}_3,b_2+c_1,b_3+c_2,\cdots,b_{2k-1}+c_{2k-2},\bar{\beta}_3+c_{2k-1},\bar{\beta}_4+\bar{\gamma}_4).
\end{equation}
If $\mathbf{b}+\mathbf{c}\neq \bar{\mathbf{b}}+\bar{\mathbf{c}}$, by (\ref{beta1}) and (\ref{gamma1}) we have that $\ell(\mathbf{b}+\mathbf{c}, \bar{\mathbf{b}}+\bar{\mathbf{c}})\geq2k$ and the conclusion follows.
If $\mathbf{b}+\mathbf{c}=\bar{\mathbf{b}}+\bar{\mathbf{c}}$, by (\ref{beta1}) and (\ref{gamma1}) we can derive that $\mathbf{b}+\mathbf{c}=(0,0,\underbrace{1,1,\cdots,1}_{2k-1},0,0)$ or $(1,1,\underbrace{0,0,\cdots,0}_{2k-1},1,1)$.

If $\mathbf{a}+\mathbf{b}=\bar{\mathbf{a}}+\bar{\mathbf{b}}$ and  $\mathbf{b}+\mathbf{c}=\bar{\mathbf{b}}+\bar{\mathbf{c}}$ at the same time, then we can derive that $\ell(\mathbf{a}+\mathbf{b},\mathbf{b}+\mathbf{c})=2k+2$ or $\ell(\mathbf{a}+\mathbf{c},\mathbf{b}+\mathbf{c})=2k$ or $\ell(\mathbf{a}+\mathbf{b},\mathbf{a}+\mathbf{c})=2k$. So, we can always obtain two distinct codewords of $\C$ such that the length of their longest common subsequence
is at least $2k$.
\EOP

\begin{lemma}\label{2n3111}
For a positive integer $k$, let $\C$ be an $[2k+3,k]$ linear insdel code over $\mathbb{F}_2$ having a codeword with $2k$ consecutive coordinates being 1. Then there exist two distinct codewords $\mathbf{u}, \mathbf{v}\in \C$ such that $\ell(\mathbf{u}, \mathbf{v})\geq 2k$.
\end{lemma}

{\it Proof:} Let $d_H$ be the minimal Hamming distance of $\C$, then there exist a codeword $\mathbf{c}\in \C$ such that $w_H(\mathbf{c})=d_H$. So, $\ell(\mathbf{c},\mathbf{0})=2k+3-d_H$. If $d_H\leq 3$, then the conclusion follows. Next, we discuss the case of $d_H> 3$.

We only prove the case that $\C$ has a codeword of the form  $\mathbf{h}=(\underbrace{1,1,\cdots,1}_{2k},0,0,0)$, and the other cases can be shown similarly.

Let $G=(G_1, G_2, \cdots, G_{2k+3})$ be a generator matrix of $\C$. Consider the linear equations
\begin{equation}\label{eq:linsys2}
(\mathbf{x}, -\mathbf{y})\underbrace{\begin{pmatrix}G_{4} & G_{5} & \cdots & G_{2k+2}\\
 G_{5} & G_{6} & \cdots & G_{2k+3}\end{pmatrix}}_M =0.
\end{equation}
The rank of $M$ is less than $2k$. So, the linear system (\ref{eq:linsys2}) has a nonzero solution $(\mathbf{x_1},-\mathbf{y_1})\in \mathbb{F}_2^{2k}$. If $\mathbf{x_1}=\mathbf{y_1}$ then we have $$(f_4(\mathbf{x_1}),f_5(\mathbf{x_1}),\cdots,f_{2k+2}(\mathbf{x_1}))=(f_5(\mathbf{y_1}),f_6(\mathbf{y_1}),\cdots,f_{2k+3}(\mathbf{y_1}))=(f_5(\mathbf{x_1}),f_6(\mathbf{x_1}),\cdots,f_{2k+3}(\mathbf{x_1})).$$ where $f_i(\mathbf{x})= \mathbf{x}G_i$ for $4\leq i\leq 2k+3$. So, $f_4(\mathbf{x_1})=f_5(\mathbf{x_1})=\cdots=f_{2k+3}(\mathbf{x_1})$. Since $d_H>3$, we derive that $\C$
has a codeword of the form $\mathbf{a}=(\ast,\ast,\ast,\underbrace{1,1,\cdots,1}_{2k})$. It is clear that $\mathbf{a}\neq \mathbf{h}$ and $\ell(\mathbf{a},\mathbf{h})\geq 2k$, then the conclusion follows. If $\mathbf{x_1}\neq\mathbf{y_1}$, and let $\mathbf{a} = \mathbf{x_1} G$ and $\bar{\mathbf{a}}= \mathbf{y_1}G$. Then $\mathbf{a}$ and $\bar{\mathbf{a}}$ have the following form:
$$\mathbf{a}=(\alpha_1,\alpha_2,\alpha_3,a_1,a_2,a_3,a_4,\cdots,a_{2k-1},\alpha_4),\,\,\,
\bar{\mathbf{a}}=(\bar{\alpha}_1,\bar{\alpha}_2,\bar{\alpha}_3,\bar{\alpha}_4,a_1,a_2,a_3,\cdots,a_{2k-2},a_{2k-1}). $$
It is obvious that the length of the longest common subsequence of $\mathbf{a}$ and $\bar{\mathbf{a}}$ is at least $2k-1$. If there exist some $i\in \{1,2,3\}$ and $j\in \{1,2,3,4\}$ such that $\alpha_i=\bar{\alpha}_j$, then $\ell(\mathbf{a},\bar{\mathbf{a}})\geq 2k$, and the conclusion follows. Otherwise, for any $i\in \{1,2,3\}$ and $j\in \{1,2,3,4\}$, we have
\begin{equation}\label{alpha2}
\alpha_i+\bar{\alpha}_j=1.
\end{equation}
By choosing proper matrices $M$ as in (\ref{eq:linsys2}), we can show that $\C$ has four codewords $\mathbf{b}$, $\bar{\mathbf{b}}$, $\mathbf{c}$ and $\bar{\mathbf{c}}$ as the following form:
\[\begin{split}
&\mathbf{b}=(\beta_1,\beta_2,b_1,b_2,b_3,b_4,\cdots,b_{2k-2},b_{2k-1},\beta_3,\beta_4),\,\, \bar{\mathbf{b}}=(\bar{\beta}_1,\bar{\beta}_2,\bar{\beta}_3,b_1,b_2,b_3,\cdots,b_{2k-3},b_{2k-2},b_{2k-1},\bar{\beta}_4);\\
&\mathbf{c}=(\gamma_1,\gamma_2,c_1,c_2,c_3,c_4,\cdots,c_{2k-2},\gamma_3,c_{2k-1},\gamma_4),\,\,\,\,
\bar{\mathbf{c}}=(\bar{\gamma}_1,\bar{\gamma}_2,\bar{\gamma}_3,c_1,c_2,c_3,\cdots,c_{2k-2},\bar{\gamma}_4,c_{2k-1}).
\end{split}\]
By analysis similar to that above, we have $\ell(\mathbf{b},\mathbf{h})\geq 2k$ if $\mathbf{b}=\bar{\mathbf{b}}$.  Otherwise, if $\mathbf{b}\neq\bar{\mathbf{b}}$, then the length of the longest common subsequence of $\mathbf{b}$ and $\bar{\mathbf{b}}$ is at least $2k-1$. Only when $\{\beta_1,\beta_2\}\cap \{ \bar{\beta}_1,\bar{\beta}_2,\bar{\beta}_3\}= \emptyset$ and $\bar{\beta}_4\notin\{ \beta_3,\beta_4\}$,
we have $\ell(\mathbf{b},\bar{\mathbf{b}})=2k-1$. In this case, for $i\in \{1, 2\}$, $j\in \{1, 2, 3\}$ and $u\in \{3, 4\}$, we have
\begin{equation}\label{beta2}
\beta_i + \bar{\beta_j} = 1 \,\, {\rm and }\,\, \beta_u + \bar{\beta}_4=1.
\end{equation}
Since $\C$ is linear, $\mathbf{a}+\mathbf{b}$ and $\bar{\mathbf{a}}+\bar{\mathbf{b}}$ are codewords of $\C$ and have the following form:
\begin{equation}\notag
\mathbf{a}+\mathbf{b}=(\alpha_1+\beta_1,\alpha_2+\beta_2,\alpha_3+b_1,a_1+b_2,a_2+b_3,\cdots,a_{2k-2}+b_{2k-1},a_{2k-1}+\beta_3,\alpha_4+\beta_4),
\end{equation}
\begin{equation}\notag
\bar{\mathbf{a}}+\bar{\mathbf{b}}=(\bar{\alpha}_1+\bar{\beta}_1,\bar{\alpha}_2+\bar{\beta}_2,\bar{\alpha}_3+\bar{\beta}_3,\bar{\alpha}_4+b_1,a_1+b_2,\cdots,a_{2k-3}+b_{2k-2},a_{2k-2}+b_{2k-1},a_{2k-1}+\bar{\beta}_4).
\end{equation}
If $\mathbf{a}+\mathbf{b}\neq\bar{\mathbf{a}}+\bar{\mathbf{b}}$, by (\ref{alpha2}) and (\ref{beta2}), we have $\ell(\mathbf{a}+\mathbf{b},\bar{\mathbf{a}}+\bar{\mathbf{b}})\geq 2k$ and the conclusion follows. If $\mathbf{a}+\mathbf{b}=\bar{\mathbf{a}}+\bar{\mathbf{b}}$, we can derive that $\mathbf{a}+\mathbf{b}=(0,0,0,\underbrace{1,1,\cdots,1}_{2k-1},0)$ or $(1,1,1,\underbrace{0,0,\cdots,0}_{2k-1},1)$.

Next, we consider the codewords $\mathbf{c}$ and $\bar{\mathbf{c}}$. If $\mathbf{c}=\bar{\mathbf{c}}$, then we derive that $\mathbf{c}=(\ast,\ast,\underbrace{1,1,\cdots,1}_{2k+1})$ or $(\ast,\ast,\underbrace{1,1,\cdots,1}_{2k-1},0,0)$ or $(1,1,\underbrace{0,0,\cdots,0}_{2k-1},1,1)$. In this case, we have $\ell(\mathbf{c},\mathbf{h})\geq 2k$ or $\ell(\mathbf{a}+\mathbf{b},\mathbf{c})\geq 2k$ or $\ell(\mathbf{a}+\mathbf{b}+\mathbf{c},\mathbf{h})\geq 2k$, then conclusion follows. If $\mathbf{c}\neq\bar{\mathbf{c}}$, it is clear that the length of the longest common subsequence of $\mathbf{c}$ and $\bar{\mathbf{c}}$ is at least $2k-1$. Only when $\{\gamma_1,\gamma_2\}\cap \{ \bar{\gamma}_1,\bar{\gamma}_2,\bar{\gamma}_3\}= \emptyset$ and $\gamma_3\neq \bar{\gamma}_4$, i.e., for $i\in \{1, 2\}$ and $j\in \{1, 2, 3\}$,
\begin{equation}\label{gamma2}
\gamma_i + \bar{\gamma_j} = 1 \,\, {\rm and }\,\, \gamma_3 + \bar{\gamma}_4=1,
\end{equation}
we have $\ell(\mathbf{c},\bar{\mathbf{c}})=2k-1$. In this case, we consider the codewords $\mathbf{a}+\mathbf{c},\bar{\mathbf{a}}+\bar{\mathbf{c}}\in \C$ as the following form:
\begin{equation}\notag
\mathbf{a}+\mathbf{c}=(\alpha_1+\gamma_1,\alpha_2+\gamma_2,\alpha_3+c_1,a_1+c_2,a_2+c_3,\cdots,a_{2k-3}+c_{2k-2},a_{2k-2}+\gamma_3,a_{2k-1}+c_{2k-1},\alpha_4+\gamma_4),
\end{equation}
\begin{equation}\notag
\bar{\mathbf{a}}+\bar{\mathbf{c}}=(\bar{\alpha}_1+\bar{\gamma}_1,\bar{\alpha}_2+\bar{\gamma}_2,\bar{\alpha}_3+\bar{\gamma}_3,\bar{\alpha}_4+c_1,a_1+c_2,\cdots,a_{2k-4}+c_{2k-3},a_{2k-3}+c_{2k-2},a_{2k-2}
+\bar{\gamma}_4,a_{2k-1}+c_{2k-1}).
\end{equation}
If $\mathbf{a}+\mathbf{c}\neq\bar{\mathbf{a}}+\bar{\mathbf{c}}$, by (\ref{alpha2}) and (\ref{gamma2}), we have $\ell(\mathbf{a}+\mathbf{c},\bar{\mathbf{a}}+\bar{\mathbf{c}})\geq 2k$, then conclusion follows.
If $\mathbf{a}+\mathbf{c}=\bar{\mathbf{a}}+\bar{\mathbf{c}}$, we have $\mathbf{a}+\mathbf{c}=(0,0,0,\underbrace{1,1,\cdots,1}_{2k-2},0,0)$ or $(1,1,1,\underbrace{0,0,\cdots,0}_{2k-2},1,1)$, then  $\ell(\mathbf{a}+\mathbf{b},\mathbf{a}+\mathbf{c})\geq2k$ or $\ell(\mathbf{b}+\mathbf{c},\mathbf{h})\geq2k$. So, we can always obtain two distinct codewords of $\C$ such that the length of their common subsequence is at least $2k$.
\EOP

By Lemma~\ref{2n3}, Lemma~\ref{2n3111} and the proof of Corollary~\ref{corhalfbound} we have the main theorem in this section.

\begin{theorem}\label{thm:optimalbound}
Let $\C$ be an $[n,k]$ linear insdel code over $\mathbb{F}_2$.
\begin{itemize}
\item[{\rm (1)}] If $n>2k$, $\mathbf{1} \notin \C$ and $\C$ is optimal with respect to the strict half-Singleton bound proposed in Theorem~\ref{thm:upperbound}, then its code length $n$ and dimension $k$ satisfy $2k+1\leq n\leq 2k+2$.
\item[{\rm (2)}] If $n\geq 2k$ and $\C$ is optimal with respect to the half-Singleton bound, then its code length $n$ and dimension $k$ satisfy $2k\leq n\leq 2k+1$.
\end{itemize}
\end{theorem}

Let $\C$ be an $[n,k]$ linear insdel code. When $k=2$, if $\mathbf{1}\notin \C$, there are 17 optimal $[5, 2]$ linear codes with respect to the strict half-Singleton bound given in Theorem~\ref{thm:upperbound}; If $\mathbf{1}\in \C$ then there are $2$ optimal $[4,2]$ linear codes  with respect to the half-Singleton bound. All these optimal linear insdel codes are listed in Table~1 and Table~2 by generators, respectively. A large number of experimental results show that Theorem~\ref{thm:optimalbound} can be strengthened into the following conjecture.

\begin{conjecture}\label{conjecture}
Let $\C$ be an $[n, k]$ linear insdel code over $\mathbb{F}_2$.
\begin{itemize}
\item[{\rm (1)}] If $n>2k$, $\mathbf{1} \notin \C$ and $\C$ is optimal with respect to the strict half-Singleton bound in Theorem~\ref{thm:upperbound}, then $\C$ has the parameters $[2k+1, k, 4]$.
\item[{\rm (2)}] If $n\geq 2k$ and $\C$ is optimal with respect to the half-Singleton bound, then $\C$ has the parameters $[2k, k, 4]$.
\end{itemize}
\end{conjecture}

\begin{table}[h]
\begin{center}
{\caption{\rm   two generators of optimal [5,2] linear insdel codes }\label{Tabvec2k+1}
\begin{tabular}{|c|c|c|}\hline
     \textbf{$\mathbf{v_1},\,\,\mathbf{v_2}$} &  \textbf{$\mathbf{v_1},\,\,\mathbf{v_2}$} &  \textbf{$\mathbf{v_1},\,\,\mathbf{v_2}$}\\\hline
  $(1,1,0,0,0),(0,0,1,1,0)$ & $(1,1,0,0,0),(0,0,1,0,1)$ & $(1,1,0,0,0),(0,0,0,1,1)$\\
  $(1,0,1,0,0),(0,0,0,1,1)$ & $(1,0,0,1,0),(0,1,1,0,0)$ & $(1,0,0,0,1),(0,1,1,0,0)$\\
  $(1,0,0,0,1),(0,1,0,1,0)$ & $(1,0,0,0,1),(0,0,1,1,0)$ & $(0,1,1,0,0),(0,0,0,1,1)$\\
  $(0,1,0,0,1),(0,0,1,1,0)$ & $(1,0,1,1,0),(0,0,1,1,1)$ & $(1,1,0,0,1),(0,0,1,1,1)$\\
  $(1,1,1,0,0),(0,0,1,1,1)$ & $(1,1,0,1,0),(0,1,0,1,1)$ & $(0,1,1,0,1),(1,0,0,1,1)$ \\
  $(1,1,1,0,0),(0,1,1,0,1)$ & $(1,1,0,0,1),(1,0,1,1,0)$ & $\quad$\\
     \hline
\end{tabular}}
\end{center}
\end{table}

\begin{table}[h]
\begin{center}
{\caption{\rm   two generators of optimal [4,2] linear insdel codes }\label{Tabvec2k}
\begin{tabular}{|c|c|} \hline
     \textbf{$\mathbf{v_1},\,\,\mathbf{v_2}$} &  \textbf{$\mathbf{v_1},\,\,\mathbf{v_2}$} \\ \hline
  $\quad\quad (1,1,0,0),\,\,\,\, (0,0,1,1)$ \quad\quad &\quad \quad $(1,0,0,1),\,\, \,\, (0,1,1,0)\quad\quad $ \\
     \hline
\end{tabular}}
\end{center}
\end{table}

\section{Strict direct upper bound}

In this section we prove the strict direct upper bound.  This bound is only true for linear insdel codes. For a linear $[n,k]$ code $\C\subset \bF_q^n$,  the subset $S\subset \{1,\cdots,n\}$ of $h$ coordinate positions is called an information free coordinate subset if  the natural projection $\Phi_S: \C \longrightarrow \bF_q^h$ defined by $\Phi_S((c_1, \cdots, c_n))=(c_{i_1},\cdots, c_{i_h})$ is surjective. It is clear $h \leq k$. When $h=k$ this is the information set. It is obvious that for any generator $k \times n$ matrix $G$ of this linear $[n, k]$ code, the columns at these positions of an information-free subset are linear independent vectors in $\bF_q^k$.

\begin{theorem}[Strict direct upper bound]\label{thm:sdub}
Let $\C \subset \bF_q^n$ be a linear $[n, k]$ code with the minimum Hamming distance $d_H$. Let ${\bf x} \in \C$ be a minimum weight codeword with its zero coordinate position set $[n]-{\rm supp}({\bf x})=\{i_1,i_2, \cdots, i_{n-d_H}\}$, where $i_1<i_2<\cdots <i_{n-d_H}$.  Suppose there are $t$ pairs of coordinate positions $\{j_u, w_u\}$, $u=1, \cdots, t$, satisfying $\{j_u, w_u\}$ is in some $[i_v+1, i_{v+1}-1]$ for $u=1, \cdots, t$, and $\{j_1, w_1, \ldots, j_t, w_t\}$ is an information-free subset. Then
$$d_I(\C) \leq 2(d_H-t).$$
\end{theorem}

{\it Proof.} Let $G$ be an $k \times n$ generator matrix of this linear code $\C$, with $n$ columns $G_1, \ldots, G_n$. Let ${\bf x}={\bf u} \cdot G$ be the minimum weight codeword claimed in the condition. Then ${\bf u}$ is a non-zero vector in $\bF_q^k$ and ${\bf u} \cdot G_j=0$ for $j=i_1, \ldots, i_{n-d_H}$. Since $\{j_1, w_1, \ldots, j_t, w_t\}$ is an information-free subset, $G_{j_1}, G_{w_1}, \ldots, G_{j_t}, G_{w_t}$ are linear independent vectors in $\bF_q^k$. Then $G_{j_1}-G_{w_1}, G_{j_2}-G_{w_2}, \ldots, G_{j_t}-G_{w_t}$ are linear independent vectors in $\bF_q^k$.  Hence we can find a non-zero vector ${\bf v} \in \bF_q^k$ satisfying ${\bf v} \cdot (G_{j_u}-G_{w_u})={\bf u} \cdot G_{w_u}$ for $1\leq u\leq t$. Now for two codewords ${\bf x}_1={\bf v} \cdot G$ and ${\bf x}_2=({\bf v}+{\bf u}) \cdot {\bf G}={\bf x}_1+{\bf x}$,  their coordinates at positions $i_1<i_2<\cdots <i_{n-d_H}$ are the same. Since
$${\bf v} \cdot G_{j_u}=({\bf v}+{\bf u}) \cdot G_{w_u},\,\, u=1, 2, \cdots, t,$$
the coordinates of ${\bf x}_1$ and ${\bf x}_2$ at position pairs $\{j_u, w_u\}$ are the same. Since $\{j_u, w_u\}$ is always in some $[i_v+1, i_{v+1}-1]$, we have a common subsequence with length  $n-d_H+t$ of these two codewords. The conclusion is proved.   \EOP

If a linear code $\C$ is projective then we have the following corollary.

\begin{corollary}\label{cor:dub}
Let $\C \subset \bF_q^n$ be a projective linear code with the minimum Hamming distance $d_H >\frac{n+1}{2}$. Then
 $$d_I(\C) \leq 2(d_H-1).$$
\end{corollary}

{\it Proof.} Since $n-d_H +1 < d_H$, for any minimum weight codeword ${\bf x}$ with zero-coordinate positions $i_1<i_2<\cdots <i_{n-d_H}$, we have an interval $[i_v+1, i_{v+1}-1]$ containing at least two support coordinate positions of ${\bf x}$. The two columns at these two positions are linear independent from the condition $d_H(\C^{\perp}) \geq 3$. The conclusion follows directly.
\EOP

Next we give two examples of linear insdel codes attaining the strict direct upper bound and an example showing that Theorem~\ref{thm:sdub} is not true for nonlinear insdel codes.

\begin{example}\label{exdub}
 Let $\C$ be an $[11,4]$ linear code over $\mathbb{F}_2$ with the following generator matrix,
\begin{equation*}
G=\left(
\begin{array}{ccccccccccc}
1 & 1 & 1 & 1 & 0 & 1 & 0 & 0 & 0 & 1 & 1\\
1 & 1 & 1 & 0 & 1 & 0 & 1 & 0 & 0 & 0 & 0\\
1 & 0 & 0 & 0 & 0 & 1 & 1 & 1 & 0 & 0 & 0\\
1 & 0 & 0 & 0 & 0 & 0 & 0 & 0 & 1 & 1 & 1
\end{array}\right)=(G_1,G_2,\cdots,G_{11}).
\end{equation*}
One can verify that $d_H(\C)=4$, and there is a minimum weight codeword $\mathbf{x}=(0,0,0,1,1,0,0,1,1,0,0)$ in $\C$ with a zero coordinate position set $\{1,2,3,6,7,10,11\}$. Then there are two pairs of coordinate positions $\{4,5\}$ and $\{8,9\}$ are in $[3,6]$ and $[7,10]$, respectively, such that $\{4,5,8,9\}$ is an information-free subset. Then by Theorem~\ref{thm:sdub}, we know that $d_I(\C)\leq 2(d_H-2)=4$. On the other hand, we can verify
that $d_I(\C)=4$. In fact, there are two distinct codewords $\mathbf{x_1}=(0,1,1,1,0,0,1,1,0,1,1)$ and $\mathbf{x_2}=\mathbf{x_1}+\mathbf{x}=(0,1,1,0,1,0,1,0,1,1,1)$ in $\C$ such that
$\ell(\mathbf{x_1},\mathbf{x_2})=9$, which is the longest common subsequence of codewords in $\C$. So, $d_I(\C)= 4$ and $\C$ attains the strict direct upper bound proposed in Theorem~\ref{thm:sdub} for $t=2$.
\end{example}

\begin{example}
Let $\C^{\prime}$ be a nonlinear code consisting of four codewords of $\C$ in Example~\ref{exdub} as follows:
\[\begin{array}{ccc}
\C^{\prime}=&\big\{ (0 , 0 , 0 , 0 , 0 , 0 , 0 , 0 , 0 , 0 , 0),\,\,(1 , 1 , 1 , 1 , 0 , 1 , 0 , 0 , 0 , 1 , 1),\\
&(1 , 0 , 0 , 0 , 0 , 0 , 0 , 0 , 1 , 1 , 1), \,\, (0 , 0 , 0 , 1 , 1 , 0 , 0 , 1 , 1 , 0 , 0)\big\}.
\end{array}\]
One can verify that $d_H(\C^{\prime})=4$ and $d_I(\C^\prime) = 8$. A minimum weight codeword $(1 , 0 , 0 , 0 , 0 , 0 , 0 , 0 , 1 , 1 , 1)$ in $\C^{\prime}$ has a zero coordinate position set $\{2,3,4,5,6,7,8\}$.
It is easy to see that there is a pair of coordinate positions $\{9, 11\}$ in $[9,11]$ such that this set is an information-free subset. However, $d_I(\C^{\prime})>2(d_H(\C^{\prime})-1)$. This example shows
that Theorem~\ref{thm:sdub} is not true for nonlinear insdel codes.
\end{example}

\begin{example}
Let $p$ be a prime number and let $e>1$ be a positive integer. Let $i_j=2^{j-1}$ for $1\leq j\leq n$ satisfying $3\cdot2^{n-2}<e$. Let $\theta$ be a primitive element in the finite field $\mathbb{F}_{p^e}$ and
\begin{equation}\notag
\C=\left\{ (\lambda+\mu\theta^{i_1},\lambda+\mu\theta^{i_2},\cdots,\lambda+\mu\theta^{i_n})\,|\, \lambda,\mu\in \mathbb{F}_{p^e} \right\}
\end{equation}
be a two-dimensional RS code of length $n$ over $\mathbb{F}_{p^e}$. Since $\C$ is an MDS code, $d_H(\C)=n-1$.
From Corollary C in \cite{Chen2022}, we know that $d_I(\C)=2n-4$. Thus, $d_I(\C)=2(d_H-1)$  and  $\C$ attains the strict direct upper bound proposed in Corollary~\ref{cor:dub}.
\end{example}

\section{Concluding remark}
In this paper, we proposed the strict half-Singleton bound for linear insdel codes without all $1$ codeword and a method to construct optimal linear insdel codes with
respect to this upper bound. Then, we proved that the length of optimal binary linear insdel codes with respect to the (strict) half-Singleton bound is about twice the dimension. A large number of experimental results suggested that optimal binary linear insdel codes have parameters $[2k, k, 4]$ or $[2k+1, k, 4]$ with respect to the half-Singleton bound or the strict half-Singleton bound proposed in Theorem~\ref{thm:upperbound}, respectively. Moreover, interestingly explicit optimal linear insdel codes attaining the (strict) half-Singleton bound, with the code length being independent of the finite field size, were obtained. Finally, we also gave the strict direct upper bound for the minimum insdel distances of linear insdel codes and optimal linear insdel codes attaining our strict direct upper bound were presented.

\end{document}